\documentclass[preprint2,11pt]{aastex61}
\received{May 2018}
\revised{}
\accepted{}
\submitjournal{ApJ}

\usepackage{hyperref}
\usepackage{amsmath}
\usepackage{amssymb}
\usepackage{graphicx}
\usepackage[utf8]{inputenc}

\DeclareRobustCommand{\ion}[2]{\textup{#1\,\textsc{\lowercase{#2}}}}
\newcommand\kms{\ensuremath{\mbox{km}\,\mbox{s}^{-1}}}
\newcommand\Teff{\ensuremath{T_\mathrm{eff}}}
\newcommand\logg{\ensuremath{\log g}}
\newcommand\feh{\ensuremath{\mbox{[Fe/H]}}}

\newcommand\vt{\ensuremath{\xi_{t}}}

\newcommand\fei{\ion{Fe}{i}}
\newcommand\feii{\ion{Fe}{ii}}

\newcommand\sii{\ion{Si}{i}}

\newcommand\he{HE\,1327$-$2326}

\newcommand\ges{G\,64$-$12}
\newcommand\bd{BD\,+44$^\circ$493}

\shorttitle{UV abundance analysis of the HMP star \he}
\shortauthors{Ezzeddine et al.}

\begin{document}
\title{Revisiting the iron abundance in the hyper iron-poor star HE~1327$-$2326 with UV COS/\textit{HST} data\footnote{Based on observations made with the NASA/ESA Hubble Space Telescope, obtained at the Space Telescope Science Institute (STScI), which is operated by the Association of Universities for Research in Astronomy, Inc. (AURA) under NASA contract NAS 5-26555. These observations are associated with program GO-14151.}}

\correspondingauthor{Rana Ezzeddine}
\email{ranae@mit.edu}

\author{Rana Ezzeddine}
\affiliation{Joint Institute for Nuclear Astrophysics - Center for Evolution of the Elements (JINA-CEE), USA}
\affiliation{Department of Physics \& Kavli Institute for Astrophysics and Space Research, Massachusetts Institute of Technology, Cambridge, MA 02139, USA}

\author{Anna Frebel}
\affiliation{Department of Physics \& Kavli Institute for Astrophysics and Space Research, Massachusetts Institute of Technology, Cambridge, MA 02139, USA}
\affiliation{Joint Institute for Nuclear Astrophysics - Center for Evolution of the Elements (JINA-CEE), USA}

\begin{abstract}
We present a new iron abundance analysis of the hyper metal-poor star \he, based on \feii\ lines detected in its UV spectral range for the first time. In a Cosmic Origins Spectrograph (COS) spectrum, five new \feii\ lines could be measured. A \sii\ line was also detected for the first time. We determine a 1D Local Thermodynamic Equilibrium (LTE) \feii\ abundance of $\mbox{[\feii/H]}=-5.99\pm0.25$. We also investigate departures from LTE for both \fei\ and \feii\ lines.  Guided by 3D Non-LTE (NLTE) analyses of other well-studied metal-poor stars, we identify potential ``residual'' 3D effects in \he\ arising from the absence of full 3D NLTE Fe calculations. 
Accordingly, we employ measurements of 10 weak \fei\ lines previously detected in an optical spectrum of \he, as no \fei\ lines are detectable in our UV spectrum. 
Following our previous work, we adopt the 1D NLTE \fei\ abundance of $\feh=-5.20\pm0.12$ for \he. 
Adopting a value based on the optical \fei\ rather than UV lines was heavily informed by our extensive investigation of model atmosphere and radiative transfer effects on different lines across the entire UV-optical wavelength range. 
An iron abundance of $\feh=-5.20\pm0.12$ is only 0.2\,dex higher than what was used in previous studies. Accordingly, no previous conclusions regarding the nature of the star are  affected.

\end{abstract}


\section{Introduction} \label{sec:intro}
The most metal-poor stars are the local equivalents to the high-redshift Universe. They retain the chemical composition of the interstellar medium at the time and place of their birth. In their atmospheres, they thus carry imprints of the nucleosynthetic signatures of their progenitor stars.
Comparing the abundance signatures of Hyper and Ultra Metal-Poor (HMP and UMP) stars (with $\mbox{[Fe/H]}\leq-5.0$ and $-5.0<\mbox{[Fe/H]}\leq-4.0$, respectively) \citep{Beers2005},
to low-metallicity supernovae yields, provide unique empirical constraints on 
the properties of their progenitors (such as masses and explosion energies), which are thought to be Population\,III first stars \citep{Frebel2015}. Also, abundance ratios such as [C/Fe] in these stars can be used to constrain 
formation scenarios of early low-mass stars \citep{frebel2007,Bennassuti2017,ji2014}. 

Galactic halo metal poor stars with $\feh <-4.0$ are however difficult to identify, but decades of searches \citep{Beers2005,Frebel2015} have delivered $\sim$\,30 such stars (e.g., \citealt{christlieb2004,caffau2012,bonifacio2015,frebel2015b,keller2014,melendez2016,aguado2018b,aguado2018a}). 

All metal-poor stars have been studied using optical spectra because spectral lines of the chemical elements of interest are found there. However, in the case of HMP and UMP stars, the absorption lines generally are weak (1-20\,m\,{\AA}), that they can only be detected in extremely high quality spectra, if at all. For iron in particular, \feii\ lines are even weaker than \fei\, lines. In the case of \he, 10 \fei\, lines had been previously detected in the optical spectrum of \he\ \citep{frebel2008} from which an iron abundance of $\mbox{[Fe/H](1D, LTE)}=-5.7\pm0.2$ was determined \citep{frebel2008}, under the assumption of 1D model atmospheres and local thermodynamic equilibrium (LTE). But no \feii\, lines could be detected. Both species together are commonly used to determine spectroscopic stellar atmospheric parameters e.g., via the ionization balance method (i.e., surface gravities, \logg), and Fe abundances, [Fe/H]. This highlights the importance of having at least 1-2 \feii\ lines available to ensure proper stellar parameter determination.

For brighter stars, such high-quality data can sometimes be obtained in reasonable amounts of observing time but in the majority of the cases, sufficient data is out of reach. However, one alternative to obtain \feii\ line measurements is from UV spectra if the stars are, once again, bright enough for such observation e.g., with the spectrographs on board the \textit{Hubble Space Telescope} (\textit{HST}). In the near-UV range, several \feii\ lines exist that are stronger than all of those potentially available in the optical regime.

HE~1327$-$2326 \citep{frebel2005} is the second brightest star ($V=13.5$) with $\feh<-4.0$ after J1808$-$5104 \citep{melendez2016} ($V=11.9$), although the former is $\sim 10$ times more metal-poor. Being one of the most iron-poor stars known to date makes it a very suitable target for studying the properties of the first stars and supernovae (SNe). We thus present the first detection of five \feii\ lines in UV spectrum of \he\, obtained with the Cosmic Origins Spectrograph (COS) on board the \textit{HST}, as well as one line for \sii. With having available \feii\ lines for the first time for this star, important constraints for the gravity,  and thus its evolutionary status can be obtained, as well as its iron abundance.

\section{Observations and data reduction}\label{Sec:obs&mes}

\begin{figure*}[ht!]
\begin{center}
\includegraphics[scale=0.35]{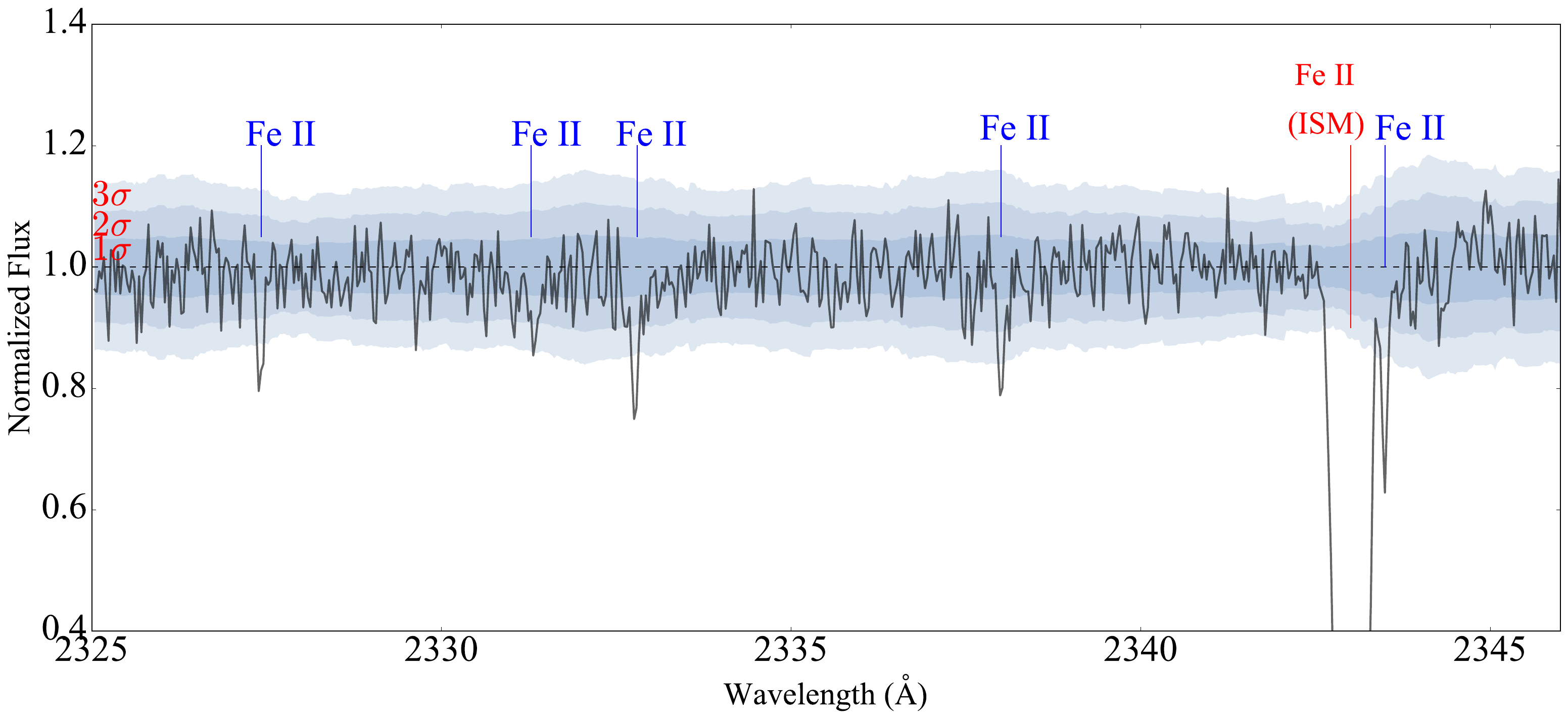}
\caption{\label{spec}Selected regions of the \he\ stacked UV spectrum where \feii\ atomic lines were detected. The different blue shaded regions represent the $1\sigma$ (blue), $2\sigma$ (light blue) and $3\sigma$ (lightest blue) detection limits, respectively, as indicated.}
\end{center}
\end{figure*}

\textit{HST}/COS observations of \he\, (Program ID: GO-14151) were obtained using the G225M grating resulting in high resolving power of $R \approx20,000$. The three UV spectral sub-ranges $\lambda2118-2151$\,{\AA}, $\lambda2216-2249$\,{\AA} and $\lambda2315-2348$\,{\AA} were covered by the data (for details of the COS instrument and performance we refer to reader to \citealt{osterman2011}). 

The total integration time was  $\sim 80.4$\,ks. 
The spectra were reduced with a custom reduction using a fixed extraction box height of 9 pixels in the cross-dispersion direction, recommended by the STScI\footnote{Space Telescope Science Institute http://www.stsci.edu/} COS instrument team \citep{snyder2017}, in order to optimize the signal-to-noise ($S/N$) ratio of the data.  
Before stacking any frames, each was individually normalized and radial velocity shifted relative to the detected \feii\ line at $\lambda 2332$\,{\AA}. The two stripes not containing this line were shifted with the same velocity as the one with the detected line because they belong to the same exposure. 
All frames for each stripe were then co-added into three final spectra. 
The final average S/N ratios are $\sim 10$\,pixel$^{-1}$ near 2120\,{\AA} and $\sim 20$\,pixel$^{-1}$ near 2350\,{\AA}. We show the part of the UV spectrum where \feii\ lines are detected in Figure\,\ref{spec}. An interstellar
\feii\ line at 2344\,{\AA} is also detected, but fortunately is sufficiently shifted away from the stellar lines. Similar ISM absorption lines such those of Na~\textsc{i} and Ca~\textsc{ii} have been previously found in the optical spectrum of \he\ \citep{frebel2005,aoki2006}.

\section{Stellar parameters}\label{Sec:stell_param}
In a previous study, \citet{frebel2005} used color-effective temperature relations from \citet{alonso1996} to determine an effective temperature for \he\, of $\Teff=6180\pm80$\,K from broadband \textit{UBVRI} photometry. 
They then used a 12 Gyr isochrone \citep{yy2002} with a metallicity of $\mbox{[Fe/H]}=-3.5$ to constrain the gravity. Two evolutionary states were possible, \logg=3.7 for a subgiant, and \logg=4.5 for a dwarf case.
To distinguish between these values, \citet{korn2009} carried out a NLTE \ion{Ca}{i}/\ion{Ca}{ii} ionization equilibrium analysis. The result favored the subgiant scenario over the dwarf case. They also argued for subgiant scenario taking into account Balmer line fitting. This was reaffirmed by \citet{mashonkina2017}, who used updated \ion{Ca}{i} atomic data in their NLTE analysis, and also found a better agreement between \ion{Ca}{i} and \ion{Ca}{ii} for \logg=3.7.
We re-investigate this issue using the recent parallax measurement ($0.8879\pm0.0235$\,mas) from the Gaia mission (DR2; \citealt{gaia2018}). We derive $\logg=3.4\pm0.3$ using fundamental relations, and adopting a stellar mass of 0.7\,M$_{\odot}$. This, once again, confirms the sub-giant scenario as the actual evolutionary status for \he.

For our new UV-based analysis of \he, we adopt \Teff=6180\,K, \logg=3.7 and a microturbulent velocity of \vt=1.7\,\kms, following \citet{frebel2005}. Accordingly, a 1D model atmospheres from \citet{castelli2004} was constructed, with input model metallicity of $\mbox{[Fe/H]}=-5.0$ 
An $\alpha$-enhancement of $\mbox{[$\alpha$/Fe]}=+0.4$ was employed throughout, following tests about the impact of the chosen $\alpha$-enhancement on the final iron abundances \citep{ezzeddine2017}.

\section{\fei\ and \feii\ abundances of \he}\label{Sec:fe_abund}

\subsection{\feii\ abundance from UV lines}\label{feii_abund}
  The equivalent widths ($EW$) of \feii\  lines in the UV spectrum were measured by convolving the COS line-spread functions \citep{ghavamian2009} with Gaussian profiles  following \citet{roederer2016} and fitting them to the lines. The uncertainties on the equivalent widths and abundance measurements were determined by altering the continuum placement and the full-width half maximum (FWHM) of the corresponding lines by $\pm 1\,\sigma$, and recording the corresponding abundance changes. The significance of the detections were assessed by dividing the equivalent width of each line by its uncertainty.
Four \feii\ lines at 2327\,{\AA}, 2332\,{\AA}, 2338\,{\AA} and 2343\,{\AA} were detected with $>3\sigma$ significance, and one \feii\ line at 2331\,{\AA} was detected with $2.1 \sigma$. The \feii\ lines as well as the $1\sigma$, $2\sigma$ and $3\sigma$ detection limits are shown in Figure\,\ref{spec}. The atomic properties of the \feii\ lines (excitation potential, oscillator strengths $\log gf$) and measured equivalent widths are listed in Table\,\ref{tab:linelist}.

\begin{deluxetable*}{c c c c c c c c }
\tablewidth{0pt}
\tabletypesize{\scriptsize}
\tablecaption{\label{tab:linelist} Atomic properties of the \feii\ absorption lines detected in the UV spectrum of \he, including the excitation potentials of their lower energy levels ($\chi$), oscillator strengths ($\log gf$), equivalent widths with their uncertainties ($EW$ and $\sigma\,EW$) as well as their detection significance (see Section\,\ref{feii_abund} for details). The last column shows our result for the determined line abundances (in LTE) at \logg=3.7.}
\tablehead{
\colhead{Absorption} & \colhead{$\lambda$} &  \colhead{$\chi$} & \colhead{$\log gf$ } & \colhead{$EW$} & \colhead{$\sigma\,EW$} & \colhead{detection} &\colhead{$\log \epsilon$(X)} \\
\colhead{line} & \colhead{{(\AA)}}  & \colhead{(eV)} &  & \colhead{(m{\AA})} & \colhead{(m{\AA})} & \colhead{significance} & \colhead{(dex)} \\}
\startdata
 \feii &        2327.39 &   0.08 &    $-0.67$ &     30.8 &  7.8  & 3.9$\sigma$ &  1.65   \\
 \feii &        2331.30 &   0.23 &    $-0.68$ &     18.0 &  8.5  & 2.1$\sigma$ &  1.43    \\
 \feii &        2332.79 &   0.04 &    $-0.19$ &     33.7 &  8.7  & 3.9$\sigma$ &  1.34    \\
 \feii &        2338.00 &   0.10 &    $-0.43$ &     32.5&  8.8  & 3.7$\sigma$ &  1.46    \\
 \feii &        2343.49 &   0.00 &    $+0.06$ &     53.8 &  10.3 & 5.2$\sigma$ &  1.56   \\
 \enddata
 \end{deluxetable*}

Line-by-line iron abundances for \he\, were then calculated using the 2017 version of the LTE radiative transfer code \texttt{MOOG}\footnote{https://www.as.utexas.edu/$\sim$chris/moog.html} \citep{sneden1973} which includes Rayleigh scattering treatment as described by \citet{sobeck2011}.
Custom spectroscopic analysis software first described in \citet{casey2014} was used. From the four \feii\ lines with $>3\sigma$ detections, we determine an average \feii\ abundance of $\mbox{[\feii/H](LTE)}=-5.99\pm0.25$ in LTE and $\mbox{[\feii/H](NLTE)}=-6.01\pm0.25$ in NLTE. The abundance uncertainties were derived following the same procedure as that used to derive the equivalent width uncertainties. The abundances (here and throughout the rest of this paper) are reported relative to the reference Fe Solar abundance $\log \epsilon(\mathrm{Fe})_{\odot}=7.50$ from \citet{asplund2009}.

\subsection{Are there systematic abundance differences from optical and UV Fe lines?}\label{Sec:iron_trend}

Using 10 \fei\, lines measured in a high resolution \textit{VLT}/UVES\footnote{The Ultraviolet and Visible Echelle Spectrograph (UVES) on the \textit{Very Large Telescope} (\textit{VLT}) at the European Southern Observatory on Cerro Paranal in the Atacama Desert.} spectrum, \citet{frebel2008} determined an \fei\ abundance of $\mbox{[\fei/H]}=-5.71\pm0.2$ for \logg=3.7. They also obtained $\mbox{[\fei/H]}=-6.01\pm0.2$ from a 3D LTE analysis, following \citep{collet2005}.  
In \citet{ezzeddine2017}, 1D LTE and 1D NLTE analyses were performed for \he\ from which $\mbox{[\fei/H](NLTE)}=-5.20\pm0.20$ and $\mbox{[\fei/H](LTE)}=-5.80\pm0.16$ were obtained, using the same lines as \citet{frebel2008}.
The LTE optical \fei\ abundance agrees within uncertainties ($\sim0.15$\,dex) with our new UV LTE \feii\ abundance, suggesting no offset as a function of a wavelength when considering LTE.

Our NLTE ``abundance correction'' obtained for \feii, defined by the difference between the average NLTE abundance and the corresponding average LTE abundance, $\Delta_{\mathrm{corr}}=\log \epsilon(\mathrm{NLTE})-\log \epsilon(\mathrm{LTE})$,  is negligible ($\Delta_{\mathrm{corr}}=-0.02$\,dex). This result is in line with previous studies \citep{mashonkina2011,bergemann2012,amarsi2016,ezzeddine2017}, as NLTE line strengths and abundances of dominant species, i.e., \feii, are not theoretically expected to significantly deviate from the LTE assumption.

Comparing this NLTE \feii\ abundance with the NLTE optical \fei\ result reveals it to be higher than that inferred from the UV \feii\, lines, by $\Delta(\log \epsilon(\fei,\mathrm{NLTE})-\log \epsilon(\feii,\mathrm{NLTE}))\\=\mathbf{0.81}$\,dex. This difference between the optical and UV abundances in NLTE is largely due to the significant deviations of the NLTE \fei\ abundance from the LTE assumption, namely by $\Delta_{\mathrm{corr}}\sim 0.6$\,dex \citep{ezzeddine2017}. Such a large discrepancy between \fei\ and \feii\ can not possibly be compensated by increasing \logg\ to attain ionization equilibrium, as it would lead to non-physical surface gravity values, i.e., $\logg\sim6$. 

\begin{figure*}[ht!]
\begin{center}
\includegraphics[scale=0.32]{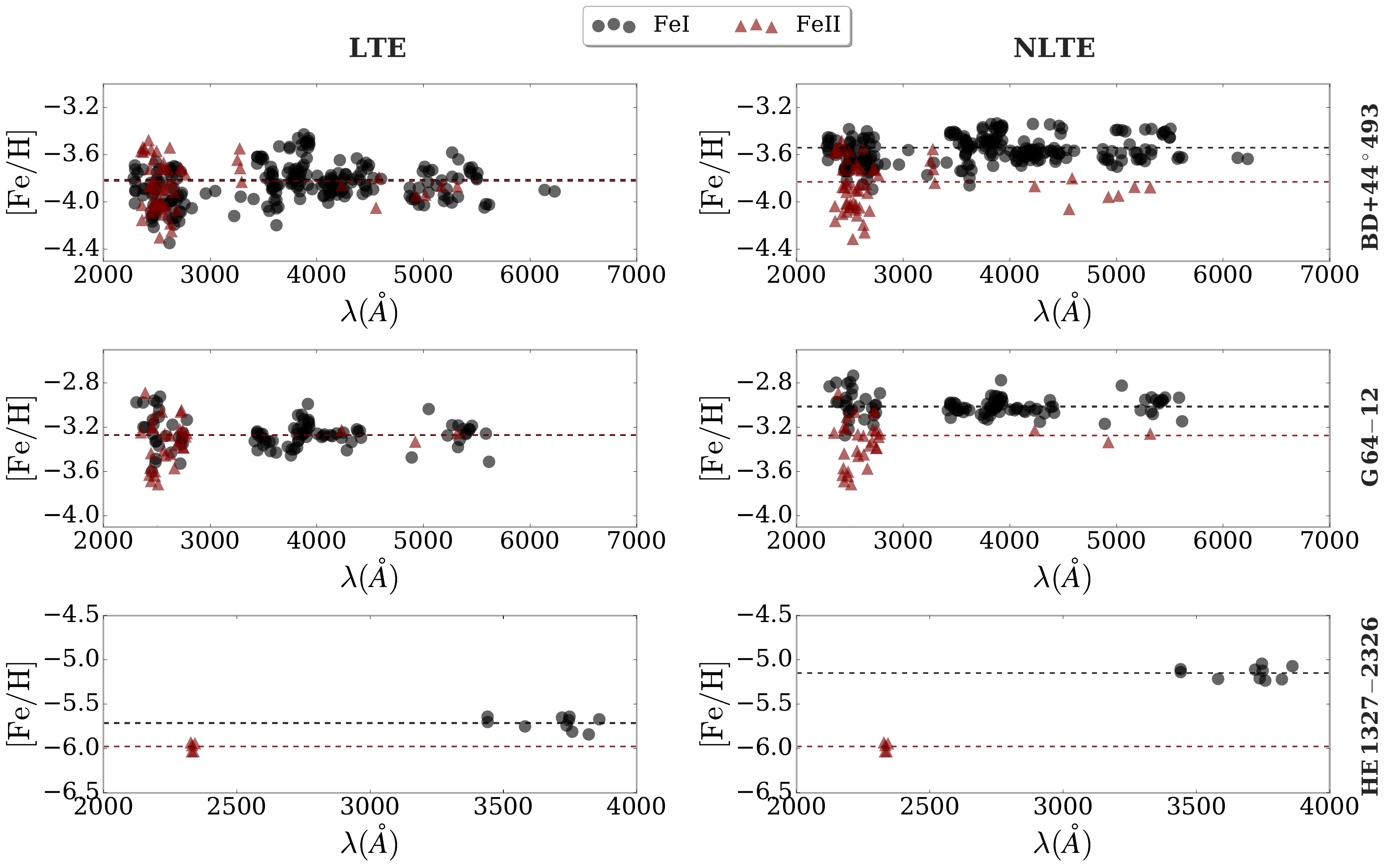}
\caption{\label{fig:fe_abund}\fei\ and \feii\ line-by-line LTE and NLTE abundances for G\,64$-$12, BD+44$^{\circ}$493 and HE\,1327$-$2327 as a function of wavelength.}
\end{center}
\end{figure*}

\subsection{Previous optical vs UV studies}

Discrepancies between the optical and UV abundances of the same species have been reported by previous studies \citep{roederer2012,lawler2013,rodererjacobson2014,roederer2014,roedererschatz2014}. 
These studies found that the abundances of individual UV lines show larger scatter than those of the optical lines (up to $\pm0.4$\,dex around the mean), and that the mean UV Fe abundances are slightly lower in the UV than in the optical (i.e., slight negative trends with decreasing wavelengths).
\citet{roederer2012}, for example, studied the LTE iron abundances in the UV and optical spectra of four reference metal-poor stars with $\mbox{[Fe/H]}>-3.1$. They found lower average abundances down to $-0.2$\,dex for lines lying between 
$\sim 3100$ and $3700$\,{\AA}, compared with those at $\lambda>4000$\,{\AA}. 

In most of these studies, however, the \fei\ lines did not continuously cover the entire wavelength range from the UV to the optical, especially toward the lowest $\lambda\sim2000$\,{\AA}, to draw firm conclusions.
It was suggested that the source of these discrepancies and decreasing abundance trends towards UV wavelengths (for the same chemical species) might be due to either NLTE effects or the underestimation of the \ion{H}{i} contribution to the continuum opacities, which becomes important toward the lower UV wavelengths at $\lambda<3500$\,{\AA}. For more details on these suggestions, we refer the reader to  Figure\,11 in \citealt{roederer2012}. However, as we show above, NLTE abundances only appear to make any discrepancies even larger.

\subsection{Investigating optical-UV abundance discrepancies with the well-studied metal-poor stars \ges\ and \bd}\label{uv-optical-diff}

We principally investigate the origin of the abundance discrepancies between lines found in the optical and UV wavelength range in \he\ and previous studies. For that, we performed a new detailed LTE and NLTE abundance analyses of two well studied extremely metal-poor stars, \ges\ and \bd. Both stars have sufficient \fei\ and \feii\ lines available across large regions of the optical and UV spectra. 
%G64-12
For \ges, we used $EW$ measurements of UV and optical Fe lines obtained from  \textit{HST}/STIS\footnote{Space Telescope Imaging Spectrograph on the \textit{HST} (Program ID GO-9049).} and \textit{VLT}/UVES spectra\footnote{Based on observations collected at the European Organisation for Astronomical Research in the Southern Hemisphere under ESO programme 67.D-0554(A) (PI = Christlieb).}, respectively \citep{roederer2018}.
We adopt $\Teff=6492$\,K and $\logg= 4.4$ as well as $\mbox{[Fe/H]} = -3.2$ and $\xi_{t}=1.5$\,km/s for input stellar parameters from \citet{roederer2018}. Their \Teff\ agrees well with the Infrared Flux Method (IRFM) \citep{casagrande2010}  value of $6464\pm150$\,K. Their surface gravity also agrees, within uncertainties, to that derived from the Gaia DR2 parallax ($3.7625\pm0.0855$\,mas) of $\logg=4.1\pm0.3$.

For \bd, we used published $EW$ measurements for UV and optical Fe lines from \citet{ito2013} and \citet{placcouv2014}, respectively.  
We adopt stellar atmospheric parameters from \citet{ito2013}, with $\Teff=5430$\,K determined from the IRFM \citep{casagrande2010}, $\logg= 3.4$, $\mbox{[Fe/H]} = -3.8$ and $\xi_{t}=1.3$\,km/s from a spectroscopic
\fei/\feii\
abundance analysis. The Gaia DR2 parallax  ($4.7595\pm0.0659$\,mas) yields a surface gravity of $\logg=3.5\pm0.3$, which is in very good agreement with the value from \citet{ito2013}.

NLTE calculations were performed following the same procedure in \citet{ezzeddine2017}, where the NLTE radiative transfer code \texttt{MULTI2.3} \citep{Carlsson1986,carlsson1992} was used with the \texttt{MARCS} \citep{gustafsson1975,gustafsson2008,plez2012} model atmospheres interpolated to the corresponding stellar parameters. 
The LTE abundances were also calculated using the same code and atmospheric models. The average abundance results obtained for \ges, \bd\ and \he\ from optical and UV \fei\ and \feii\ lines (whenever possible) are displayed in Table\,\ref{tab:benchmak_tests}. Line-by-line abundances as a function of wavelengths are also shown for the three stars in Figure\,\ref{fig:fe_abund}. 

In LTE, no systematic differences are found between the abundances derived from the UV and optical lines or those derived from \fei\ and \feii\ lines for \bd, \ges\ and \he. Any small differences are within the abundance uncertainties of the three stars. In NLTE, however, the abundances obtained from  both optical and UV \fei\ lines are systematically higher than those from the \feii\ lines from either region (when available). The \fei-\feii\ discrepancies are $+0.30$\,dex for \bd, $+0.33$\,dex for \ges\ and $+0.80$\,dex for \he. It is important to note, though, that the abundances inferred from UV \feii\ lines show larger scatter ($\pm 0.2$\,dex for \ges\ and \bd) around the mean values as compared to those inferred from optical \feii\ lines ($\pm0.14$\,dex for \bd\ and $\pm0.05$\,dex for \ges), or UV and optical \fei\ lines ($\sim \pm0.15$\,dex). The same large scatter is obtained in both LTE and NLTE analyses. This is in line with previous LTE Fe abundance studies based on UV lines \citep[e.g.,][]{roederer2012}.

\begin{deluxetable*}{l c c c c c c c c c c c c c}
\tablewidth{0pt}
\tabletypesize{\scriptsize}
\tablecaption{\label{tab:benchmak_tests} LTE and NLTE average line \fei\ and \feii\ abundances of the test comparison stars \bd\ and \ges\ as well as \he, determined from their UV and optical spectra respectively.}
\tablehead{
&\multicolumn{2}{c}{\bd}  &\multicolumn{2}{c}{\ges}& \multicolumn{2}{c}{\he}\\
&\colhead{LTE} & \colhead{NLTE} &   \colhead{LTE} & \colhead{NLTE} &   \colhead{LTE} & \colhead{NLTE}}
\startdata
$\mbox{[\fei/H]}^{\mathrm{Op}}$ & $-3.81\pm0.15$ & $-3.54\pm0.10$ &  $-3.27\pm0.10$ & $-3.02\pm0.10$ &  $-5.80\pm0.16$ & $-5.20\pm0.12$ \\
$\mbox{[\feii/H]}^{\mathrm{Op}}$ & $-3.83\pm0.14$ & $-3.83\pm0.14$ &  $-3.27\pm0.05$ & $-3.27\pm0.05$ &  \nodata & \nodata\\
$\mbox{[\fei/H]}^{\mathrm{UV}}$ & $-3.95\pm0.15$ & $-3.61\pm0.13$ &  $-3.24\pm0.18$ & $-2.99\pm0.13$ &  \nodata & \nodata \\
$\mbox{[\feii/H]}^{\mathrm{UV}}$ & $-3.86\pm0.20$ & $-3.87\pm0.21$ &  $-3.34\pm0.20$ & $-3.34\pm0.21$ &  $-5.99\pm0.25$ & $-6.01\pm0.25$ \\ 
$\mbox{[\fei/H]}^{\mathrm{Op+UV}}$ & $-3.85\pm0.16$ &  $-3.56\pm0.11$ &  $-3.26\pm0.16$ & $-3.01\pm0.10$ &  $-5.80\pm0.16$ & $-5.20\pm0.12$ \\ 
$\mbox{[\feii/H]}^{\mathrm{Op+UV}}$ & $-3.86\pm0.19$ & $-3.86\pm0.19$ &  $-3.33\pm0.15$ & $-3.34\pm0.12$ &  $-5.99\pm0.26$ & $-6.01\pm0.25$\\
\enddata
\end{deluxetable*}

In order to further understand this discrepancy, we considered other studies on e.g. \ges.
\citet{amarsi2016} performed a full 3D NLTE abundance analysis for \ges\ and three other well-studied metal-poor stars. As part of their work, they found their 1D NLTE  abundances for \ges\ to be higher than the 1D LTE values, up to $0.23$\,dex for \fei\ but are, of course, negligible for \feii. This is very similar to our findings ($\Delta_{\mathrm{corr}}=0.25$\,dex.), albeit using a slightly lower \logg\ value by 0.14\,dex than ours.

Also, \citet{amarsi2016} compared 3D NLTE and 1D NLTE abundances in \ges. They found non-negligible positive abundance corrections of $+0.16$\,dex for \feii\ and $\sim+0.11$\,dex for \fei\ due to 3D effects in \ges. Interestingly, both these 3D NLTE \fei\ and \feii\ abundance are increased (although not at the same level) compared to the 1D NLTE case. This differential increase of both \fei\ and \feii\ abundances brings them into agreement with each other, to within $\sim 0.1$\,dex, when abundance uncertainties are also $\sim 0.1$\,dex. 

The abundances derived in this combined 3D NLTE framework furthermore suggests \fei\ and \feii\ corrections to increase even more toward lower metallicities relative to just 1D NLTE \citep{amarsi2016,nordlander2017}. This behavior has already been documented as part of the full 3D NLTE analysis of the most-iron poor giant star SMSS~0313$-$6708 with [Fe/H]$<-7.0$ \citep{nordlander2017}. ``Residual'' 3D corrections of $\Delta^{\mathrm{3D}}=\log \epsilon(\mathrm{3D,NLTE}) -\log\epsilon(\mathrm{1D,NLTE}) =0.2$\,dex were determined for a putative \fei\ line in this star. 
Moreover, the corresponding residual 3D correction for \feii\ lines in SMSS~0313$-$6708 is expected to be larger than that for \fei\ \citep{amarsi2016}, although, unfortunately, no explicit calculations were reported for SMSS~0313$-$6708. In addition, they report a 1D NLTE effect of $\Delta_{\mathrm{corr}}=0.6$ (compared to 1D LTE). This is larger than the residual 3D correction, but comparable to what we find for \he\ in the present study. Finally, \citet{nordlander2017} found that the 3D LTE framework can heavily underestimate the \fei\ abundances by $\sim 1$\,dex compared to the full 3D NLTE case.

Consequently, they deduced that the 3D NLTE \fei\ abundance ($\mbox{[Fe/H]}<-6.53\pm0.5$) is the best avenue for determining the iron abundance of this star, but if a 3D analysis is not possible, the 1D NLTE case 
($\mbox{[Fe/H]}<-6.73\pm0.2$) provides a robust estimate of the Fe abundance instead. This will be the case for most analyses, given that full 3D NLTE calculations are based on complex and lengthy computational requirements.

By analogy, we thus conclude that the discrepancies observed in \he\ between the UV \feii\ line and the optical \fei\ line abundances may well be due to these unaccounted 3D residual corrections. 
Further exploration of the issue in terms of 3D modeling is warranted, but beyond the scope of the present study.

\subsection{Adopted iron abundance of \he}

Our investigation into understanding the differences obtained between the abundances of \fei\ and \feii\ arising from LTE, NLTE and 3D calculations for \ges, \bd\ and SMSS~0313$-$6708 yielded that the 1D NLTE \fei\ abundance is presently the most robust indicator to use, if a full 3D NLTE calculation is not available \citep{amarsi2016,nordlander2017}.

Guided by these results, for \he, we thus adopt the 1D NLTE abundance of $\feh=-5.20\pm0.12$ inferred from optical \fei\ lines as our final [Fe/H] value. This new result is slightly higher than what was previously adopted $\feh=-5.4$ \citep{frebel2005}, which was an LTE value corrected with an arbitrarily chosen NLTE correction of 0.2\,dex. A value of $\feh=-5.2$ should be more physically meaningful for any interpretation of the origin of this star and its abundance pattern than any other values. Performing a computationally challenging full 3D NLTE abundance analysis in the future would hopefully show agreement between the abundances of both \fei\ and \feii\ species, within uncertainties. 

\section{Silicon UV abundance}
In addition to the \feii\ lines, we also detected for the first time a \sii\ line at 2124.12\,{\AA} 
in the UV spectrum of \he. From that, using the same procedure as for \feii\ reported in Section\,\ref{Sec:fe_abund}, we determine a Si abundance of $\mbox{[Si/H]}=-4.51\pm0.23$.

The UV Si abundance is higher than the optical upper limit reported by \citet{ji2014} of $\mbox{[Si/H]}=-5.40$. These authors investigated the role of critical gas fragmentation in first low-mass stars formation and compared their models to abundance measurements of Si in UMP stars. 
Their upper limit falls right below that of the supernova shock dust size distribution limit (see Figure\,5 in \citealt{ji2014}). The implication was previously that \he\ could not have formed from gas solely cooled by silicon-based dust, but instead by some other mechanism that enables low-mass star formation. 

At face value, our new $\mbox{[Si/H]}$ measurement only places \he\ below the standard shock dust size distribution limit which is higher than the supernova shock dust distribution. Assuming that no discrepancies exist between the Si abundances inferred from UV and optical lines, this would suggest that the formation \he\ (and perhaps that of other UMP stars) from early gas clouds could have been driven by dust-induced gas cooling although the high carbon abundance of 
\he\ still points to a formation from fine-structure line cooled gas \citep{frebel2007}.

\section{Summary}\label{Sec:conc}
We analyzed the UV spectrum of \he\ between 2120 and 2360\,{\AA}, in which 5 new \feii\ lines as well as one \sii\ line have been detected. We obtain a 1D LTE \feii\ abundance of [\feii/H]=$-5.99\pm0.25$ from the four \feii\ lines with $>3\sigma$ detection significance. This value is $\sim0.8$\,dex lower than the 1D NLTE iron abundance inferred from optical \fei\ lines of $\feh=-5.20\pm0.12$. We investigated  previous claims that discrepancies exist between abundances inferred from UV and optical lines using the two well-studied metal-poor stars \bd\ and \ges. We find that no systematic differences are found in LTE when sufficient UV and optical lines are used in the analysis, however, \feii\ abundances of UV lines display larger scatter from the mean values than their optical counterparts.
On the other hand, non-negligible differences are found between abundances inferred from \fei\ and \feii\ lines in NLTE in both stars, as is the case for \he. These differences could be attributed to ``residual'' 3D effects, as discussed in Section\,\ref{uv-optical-diff}.

We therefore adopt the 1D NLTE iron abundance of $\feh=-5.20\pm0.12$ as the final iron abundance of \he. This value can be considered as best possible if a full 3D NLTE calculation is lacking. This follows recommendations by \citet{nordlander2017} who performed both 1D and 3D NLTE analyses in an HMP star.

This iron abundance in \he\ is just 0.2\,dex higher than of $\mbox{[Fe/H]}=-5.40\pm0.2$ adopted by \citet{frebel2005}. They used this latter value in comparing the overall elemental abundance ratio pattern of \he\ (relative to Fe) to supernovae explosion nucleosynthesis yields of Population\,III stars, in order to derive its progenitor star properties. Such a negligible difference between our values, relative to the full abundance scale spanning over 6\,dex (See Figure\,2 in \citealt{frebel2005}), does not alter the conclusions deduced by their results, which are still expected to hold.

\acknowledgements
We thank the anonymous referee for their time and useful comments which helped improve this manuscript. We thank Ian U. Roederer for the useful discussion and advice on this work, and for providing the equivalent widths and stellar parameters measurements of \ges\ as well as the equivalent widths for \he.
R.E. acknowledges support from a JINA-CEE fellowship (Joint Institute for Nuclear Astrophysics - Center for the Evolution of the Elements), funded in part by the National Science Foundation under Grant No. PHY-1430152 (JINA-CEE). A.F. is supported by NSF-CAREER grant AST-1255160. AF acknowledges support from the Silverman (1968) Family Career Development Professorship. We also
thank and appreciate the expert assistance of Jason Tumlinson in the custom extraction of the UV spectra. This research has made use of NASA’s Astrophysics Data System Bibliographic Services, the arXiv pre-print server operated by Cornell University and the Atomic Spectra Database hosted by the National Institute of Standards and Technology (NIST).

\software{MOOG \citep{sneden1973,sobeck2011},  MULTI2.3 \citep{Carlsson1986,carlsson1992}, MARCS \citep{gustafsson1975,gustafsson2008,plez2012}}

\bibliography{ref}

\clearpage

\end{document}